\documentclass[a4]{article}
\usepackage{amsmath,amssymb} %for dealing with mathematics,
\usepackage{graphicx} %for dealing with figures.

\begin{document}

\begin{center}
{\LARGE Noncommutative Spacetime Realized in} \\
{\LARGE  $AdS_{n+1}$ Space }  \vspace{2mm}\\
{\large  Non-Local Field Theory out of Noncommutative Spacetime}\vspace{5mm}

S. Naka${}^1$, H. Toyoda${}^2$, T. Takanashi${}^3$, and E.Umezawa${}^4$ \vspace{3mm}\\

{\it
${}^1$ Department of Physics, College of Science and Technology Nihon University, Tokyo 101-8308, Japan \\
${}^2$ Junior College, Funabashi Campus, Nihon University, Funabashi 274-8501, Japan \\
${}^3$ Ibaraki Prefectural University of Health Sciences, 4669-2 Ami, Ami-machi, Inashiki-gun, Ibaraki 300-0394, Japan \\
${}^4$ School of Health Sciences, Fujita Health University, Aichi 470-1192, Japan}

\end{center}

\begin{abstract}%

In $\kappa$-Minkowski spacetime, the coordinates are Lie algebraic elements such that time and space coordinates do not commute, whereas space coordinates commute each other. The non-commutativity is proportional to a Planck-length-scale constant $\kappa^{-1}$, which is a universal constant other than the light velocity under the $\kappa$-Poincare transformation. In this sense, the spacetime has a structure called as \lq\lq Doubly Special Relativity\rq\rq . Such a noncommutative structure is known to be realized by $SO(1,4)$ generators in 4-dimensional de Sitter space. In this paper, we try to construct a nonommutative spacetime having commutative n-dimensional Minkowski spacetime based on $AdS_{n+1}$ space with $SO(2,n)$ symmetry. We also study an invariant wave equation corresponding to the first Casimir invariant of this symmetry as a non-local field equation expected to yield finite loop amplitudes.
\end{abstract}

\section{Introduction}

The $\kappa$-Minkowski spacetime is a noncommutative spacetime characterized by an algebraic structure with a constant $\kappa$ other than the light velocity; in this sense, the framework of $\kappa$-Minkowski spacetime is called as doubly special relativity (DSR)\cite{kappa-Minkowski}. The $\kappa$ is a Planck energy scale constant, which is usually said to be a trace of quantum gravity through the combination $\kappa \sim \sqrt{G/\Lambda}$ (or $\sqrt{\hbar/G}$) in some limit of $G,\Lambda \rightarrow \infty$\cite{Amelino-Camelia}. Here, $G$ and $\Lambda$ are respectively the gravitational constant and the cosmological constant.

Associated with this dimensional constant, the coordinates of the $\kappa$-Minkowski spacetime form the Lie algebra characterized by 
\begin{align}
 [\hat{x}_{i},\hat{x}_{j}] &=0 , \label{al-01}\\
 [\hat{x}_{0},\hat{x}_{i}] &=-i\kappa^{-1}\hat{x}_{i} , \label{al-02}
\end{align}
where $i$ runs over $(1,2,3)$. The characteristic relations (\ref{al-01}) and (\ref{al-02}) spoil the symmetry under the Lorentz boost in the Minkowski spacetime, although they are symmetric under the space rotation. Nevertheless, this framework is symmetric under the $\kappa$-Poincare transformation, which reduces to the Lorentz transformation according as $\kappa \rightarrow \infty$.

Historically, another Lie algebra type of noncommutative spacetime was first discussed in 1947 by H. S. Snyder\cite{Snyder} in the context of non-local field theory extended in spacetime with the fundamental length $l\sim\kappa^{-1}$. Although the algebraic structure of Snyder's spacetime is slightly different from that of the $\kappa$ Minkowski spacetime, the symmetry under the Lorentz boost is broken in this spacetime too. In both types of noncommutative space times, the dispersion relation of particles embedded in this spacetime becomes highly non-linear due to $\kappa\neq 0$. As a result of those non-linear dispersion relations, the wave equations of particles possess non-local structure by regarding those equations as effective ones in the usual commutative spacetime.

On the other hand, it is well known that H. Yukawa proposed an attempt of non-local field theory\cite{Yukawa}, the bi-local field theory, in the same period as Snyder's noncommutative spacetime theory appeared. Yukawa's attempt is motivated by a unified description of elementary particles and to get divergence free field theories by introducing a fundamental length in spacetime. After long history according to this line of thought, Yukawa arrived at an idea of field theory such as elementary domains\cite{domain}, which obeys a difference equation instead of a differential equation. The equation of domain keeps the Lorentz invariance, but it is not consistent with the causality, since the field equation allows timelike extension of fields.  

Although the field theories based on $\kappa$-Minkowski spacetime and domain like non-local field theories are standing on different bases of thought, they look like to have close connection each other. The purpose of this paper is, thus,  to study the relationship between a domain type of field theory and a non-local field theory based on a $\kappa$-Minkowski like spacetime, which is modified so as to be symmetric under the Lorentz boost.

In the next section, we formulate the $\kappa$-Minkowski spacetime from the viewpoint regarding noncommutative coordinates as $SO(1,4)$ generators in $dS_4$ space, the 4-dimensional de Sitter space. As an extension of this formulation, in section 3, we discuss a modified non commutative spacetime realized in $AdS_{n+1}$ space so that the framework is symmetric under the Lorentz transformation. I section 4, we discuss a wave equation of a non-local field characterized by the first Casimir invariant in this space; and, a detailed analysis of on-mass-shell particles is given. In section 5,  discussions on $\phi^3$ type of interaction of such a field are made. Therein, we show that some loop diagrams become finite due to a non-unitary structure of such a field in the energy scale of $\kappa$ .

Section 6 is the discussion and summary. In appendices A and B, the mathematical background of \S 2 and \S 3 is summarized.
\section{$\kappa$-Minkowski spacetime based on $dS_4$}

A simple way to construct the Lie algebraic coordinates (\ref{al-01}) and (\ref{al-02}) is to start from $dS_4$ with coordinates $y=(y^A)=(y^0,y^i,y^4),(i=1,2,3)$ characterized by

\begin{equation}
 y^2=g_{AB}y^A y^B=(y^0)^2-(y^1)^2-(y^2)^2-(y^3)^2-(y^4)^2=-\kappa^2.
\end{equation}
In terms of these coordinates, the generators of $SO(1,4)$ isometric group can be written as
\begin{equation}
M_{AB} = i\kappa^{-1}(y_A\partial_B-y_B\partial_A),~(\partial_A=\frac{\partial}{\partial y^A}),
\end{equation}
to which one can verify that
\begin{equation}
  [M_{AB},M_{CD}] =i\kappa^{-1} (g_{BC}M_{AD}+g_{AD}M_{BC}-g_{AC}M_{BD}-g_{BD}M_{AC} ). \label{SO(1,4)}
\end{equation}
It is obvious that there are ten independent components of these generators; that is, the generators of space rotation $\{M_{ij}\}$, the Lorentz boost $\{M_{i0}\}$, and remaining four generators $\{M_{\mu 4}\},((\mu)=(0,i))$. The noncommutative coordinates $\hat{x}^0$ and $\hat{x}^i$ are constructed out of those remaining generators by
\begin{align}
 \hat{x}_0 &= M_{04}, \\
 \hat{x}_i &= M_{i0}+ M_{i4}.
\end{align}
Here, the inverse sign $g_{00}=-g_{44}$ in the metric is essential to realize the commutation relations (\ref{al-01}) and (\ref{al-02})
\footnote{
By definition, $\hat{x}^i$ is a three vector under the rotation by $\{M_{ij}\}$; but it is not any three vector under the Lorentz boost by $\{M_{i0}\}$. To make clear the meaning of $\hat{x}^\mu$, let us put $y_4=+\sqrt{y_\mu y^\mu+\kappa^2}$ after calculating the commutation relations $[\hat{x}_\mu,y_\nu]$, then we obtain
\begin{align*}
  [\hat{x}_0,y_0] &=-i\kappa^{-1}y_4\rightarrow -i,~(\kappa \rightarrow \infty) ,  \\
 [\hat{x}_i,y_j] &=i\kappa^{-1}g_{ij}(-y_0-y_4)\rightarrow -ig_{ij},~(\kappa \rightarrow \infty). 
\end{align*}
Thus, $(\hat{x}^\mu)$ tends to the $p$-representation of coordinates in flat Minkowski spacetime in the limit $\kappa\rightarrow \infty$. In other words, $\{y^\mu\}$ is the momentum space in flat Minkowski spacetime.
 }.
For the latter purpose, it is convenient to introduce light-cone variables between $(y^0,y^4)$; that is, we put $(y^A)=(y^i,y^{+},y^{-}),(A=i,\pm)$, where $y^{\pm}=y^0\pm y^4$. Then, we can write the invariant length in $dS_4$ as $y^2=\bar{g}_{AB}y^Ay^B=-(y^i)^2+y^{+}y^{-}$ with the metric
\begin{equation}
(\bar{g}_{AB})=
\begin{pmatrix}
-1 & 0 & 0 & 0 & 0 \\
0 & -1 & 0 & 0 & 0 \\
0 & 0 & -1 & 0 & 0 \\
0 & 0 & 0 & 0 & \frac{1}{2} \\
0 & 0 & 0 & \frac{1}{2} & 0 
\end{pmatrix}.
\end{equation}
In this basis, the noncommutative coordinates can be expressed as
\begin{align}
 \hat{x}_i &= 2i\kappa^{-1}(y_i\partial_{+}-y_{+}\partial_i)=2M_{i+}, \\
 \hat{x}_0 &= 2i\kappa^{-1}(y_{-}\partial_{+}-y_{+}\partial_{-})=2M_{-+}.
\end{align} 

In order to find the invariant wave equation in the $\kappa$-Minkowski spacetime, let us consider the unitary transformation $U(\omega)=e^{\frac{i}{2}\omega^{AB}M_{AB}}$, which causes a finite $SO(1,4)$ transformation in $dS_4$ in such a way that $y^A=U(\omega)y^AU^\dag(\omega)={(e^{\kappa^{-1}\boldsymbol{\omega}})^A}_By^B$ with $\boldsymbol{\omega}=(\omega^A{}_B)$. In particular, for the contracted transformation defined by $\omega^{BC}=a^Bb^C-b^Ba^C$ with $(b^i,b^{-},b^{+})=(0,2,0)$, the exponent of $U(\omega)$ reduces to $\frac{1}{2}\omega^{BC}M_{BC}=a^{-}\hat{x}^0-a^i\hat{x}^i$; namely, $U(\omega)$ becomes an exponential function in the $\kappa$-Minkowski spacetime in this contracted case. Furthermore, $a^i$ and $a^{-}$ are related to four momenta conjugate to $\{\hat{x}^\mu\}$ in some sense, which will be discussed soon.

Now, $SO(1,4)$ transformation of a c-number vector $u=(u^A),(u^2=-\kappa^2)$ in $dS_4$ is defined by $U(\omega)(u\cdot y)U(\omega)^\dag=\{(e^{-\kappa^{-1}\boldsymbol{\omega}})_A{}^B u^A\}y_B=u(\omega)\cdot y$, so that $u(\omega)^2=-\kappa^2$ holds. If we choose, for sake of simplicity, $u=(u^0,u^i,u^4)=(0,0,0,0,\kappa)$, then $u(\omega)^A=\kappa(e^{-\kappa^{-1}\boldsymbol{\omega}})^A{}_4$ becomes a non-linear realization of a vector in $dS_4$ in terms of $(a^i,a^{-})$. Here, a little calculation leads to the explicit form of $e^{-\kappa^{-1}\boldsymbol{\omega}}$ such that (Appendix A)
\begin{equation}
 e^{-\kappa^{-1}{\boldsymbol{\omega}}}={\boldsymbol{1}}+\frac{1}{(a^{-})^2}\left\{\cosh (\kappa^{-1}a^{-})-1\right\}{\boldsymbol{\omega}}^2-\frac{1}{a^{-}}\sinh(\kappa^{-1}a^{-}){\boldsymbol{\omega}}.
\end{equation}
By taking $(\boldsymbol{\omega}_{04},\boldsymbol{\omega}_{i4},\boldsymbol{\omega}_{44})=(-a^{-},-a_i,0)$ and $(\boldsymbol{\omega}^2_{04},\boldsymbol{\omega}^2_{i4},\boldsymbol{\omega}^2_{44})=(-\boldsymbol{a}^2,-a^{-}a_i,-(a^{-})^2+\boldsymbol{a}^2)$ into account, we thus arrive at the expression
\begin{align}
 \tilde{u}_i(\omega) &=(1-e^{-\kappa^{-1}a^{-}})\frac{a_i}{a^{-}} \\
 \tilde{u}_0(\omega) &=-\left\{\cosh(\kappa^{-1}a^{-})-1 \right\}\left(\frac{\boldsymbol{a}}{a^{-}}\right)^2+\sinh(\kappa^{-1}a^{-}), \\
 \tilde{u}_4(\omega) &=-\cosh(\kappa^{-1}a^{-})+\left\{\cosh(\kappa^{-1}a^{-})-1 \right\}\left(\frac{\boldsymbol{a}}{a^{-}}\right)^2,
\end{align}
where we have written $\tilde{u}_A(\omega)=\kappa^{-1}u(\omega)_A$; henceforth, we use the same notation $\tilde{f}=\kappa^{-1}f$ for any $f$. Further, we note that the vector $(\tilde{u}^A)$ is usually introduced in relation to the bicovariant differentials of $U(\omega)$, since  $\tilde{u}(\omega)^A=(e^{-\kappa\boldsymbol{\omega}})^A{}_4$ satisfies (Appendix A)
\begin{align}
 \mbox{d}U(\omega) =i\kappa\left\{dx_\mu\tilde{u}^\mu +dx_4(\tilde{u}^4 -1)\right\}U(\omega). \label{bicovariant}
\end{align}

 The next task is to identify $U(\omega)$ to an ordered exponential function $e^{-ik^0\hat{x}^0+ik^i\hat{x}^i}$ by some way: the typical cases are
\begin{align}
 \hat{e}(k) &= e^{-i(k^0\hat{x}^0-k^i\hat{x}^i)}=e^{-ik^0\hat{x}^0}e^{i\frac{e^{\tilde{k}^0}-1}{\tilde{k}^0}k^i\hat{x}^i}=e^{i\frac{1-e^{-\tilde{k}^0}}{\tilde{k}^0}k^i\hat{x}^i}e^{-ik^0\hat{x}^0} , \\
 \hat{e}_R(k) &= e^{ik^i\hat{x}^i}e^{-ik^0\hat{x}^0}=e^{-i(k_R^0\hat{x}^0-k_R^i\hat{x}^i)} , \\
 \hat{e}_L(k) &=e^{-ik^0\hat{x}^0}e^{ik^i\hat{x}^i}=e^{-i(k_L^0\hat{x}^0-k_L^i\hat{x}^i)}, \\
 \hat{e}_S(k) &= e^{-\frac{i}{2}k^0\hat{x}^0}e^{ik^i\hat{x}^i}e^{-\frac{i}{2}k^0\hat{x}^0}=e^{-i(k_S^0\hat{x}^0-k_S^i\hat{x}^i)}, 
\end{align}
where
\begin{align}
 k_R=(k_R^0,k_R^i) &=(k^0,\frac{\tilde{k}^0}{1-e^{-\tilde{k}^0}}k^i) , \\
 k_L=(k_L^0,k_L^i) &=(k^0,\frac{\tilde{k}^0}{e^{\tilde{k}^0}-1}k^i) , \\
 k_S=(k_S^0,k_S^i) &=(k^0,\frac{\tilde{k}^0}{e^{\tilde{k}^0/2}-e^{-\tilde{k}^0/2}}k^i).
\end{align}
For example, if we put $U(\omega)=\hat{e}_R(k)$, the right ordering of exponential function , we get the well known expression of the vector in $dS_4$ such that
\footnote{
Similarly, the substitutions $(a^{-},a^i)\rightarrow -k_{L/S}$ give yields the other expressions of vectors in $dS_4$: \vspace{1mm}\\
\hspace{15mm}
\begin{minipage}[t]{4cm}
\underline{~$(a^{-},a^i)\rightarrow -k_{L}$~}
\begin{align*}
 \tilde{u}^i(k)_L &=-\tilde{k}^i , \\
 \tilde{u}^0(k)_L &=-\frac{1}{2}e^{-\tilde{k}^0}\tilde{\boldsymbol{k}}^2-\sinh(\tilde{k}^0) , \\
 \tilde{u}^4(k)_L &=\cosh(\tilde{k}^0)-\frac{1}{2}e^{-\tilde{k}^0}\tilde{\boldsymbol{k}}^2 .
\end{align*}
\end{minipage}
\hspace{10mm}
\begin{minipage}[t]{4cm}
\underline{~$(a^{-},a^i)\rightarrow -k_{S}$~}
\begin{align*}
 \tilde{u}^i(k)_S &=-e^{\frac{\tilde{k}}{2}}\tilde{k}^i , \\
 \tilde{u}^0(k)_S &=-\frac{1}{2}\tilde{\boldsymbol{k}}^2-\sinh(\tilde{k}^0) , \\
 \tilde{u}^4(k)_S &=\cosh(\tilde{k}^0)-\frac{1}{2}\tilde{\boldsymbol{k}}^2 .
\end{align*} 
\end{minipage}\vspace{-5mm}
}
\begin{align}
 \tilde{u}^i(k)_R &=-e^{\tilde{k}^0}\tilde{k}^i , \label{ui}\\
 \tilde{u}^0(k)_R &=-\frac{1}{2}e^{\tilde{k}^0}\tilde{\boldsymbol{k}}^2-\sinh(\tilde{k}^0) , \label{u0}\\
 \tilde{u}^4(k)_R &=\cosh(\tilde{k}^0)-\frac{1}{2}e^{\tilde{k}^0}\tilde{\boldsymbol{k}}^2 .
\end{align}
Therefore, if we put $P^A(k)=-\kappa\tilde{u}^A(k)_R$, then eq.(\ref{bicovariant}) can be read as 
\begin{equation}
 d\hat{e}_R(k)=-i\left\{dx_\mu P^\mu(k) +dx_4(P^4(k) -\kappa)\right\}\hat{e}_R(k) ,
\end{equation}
where $P(k)^A,(A=\mu,4)$ are five momenta satisfying $P(k)^A P(k)_A=-\kappa^2$.

Finally, we discuss the $SO(1,3)$ transformation realized in $\{\tilde{u}(k)^\mu\}$ through the transformation of $\{k^\mu\}$; then, their resultant form should be 
\begin{equation}
 \hat{\mathcal L}_{\mu\nu}\tilde{u}_{\rho}=(g_{\mu\rho}\tilde{u}_{\nu}-g_{\nu\rho}\tilde{u}_{\mu}),~~~\hat{\mathcal L}_{\mu\nu}\tilde{u}_{4}=0,
\label{SO(1,3)}
\end{equation}
so that the constraint $\tilde{u}^A\tilde{u}_B=-1$ holds. The $\hat{\mathcal L}_{\mu\nu}$'s are actions on $k^\mu$ causing non-linear transformations in general. The space rotation is simple, since $\tilde{u}(k)^0$ and $\tilde{u}(k)^i$ transform respectively as a scalar and a vector under the rotation of $\{k^i\}$; then, the action of $\hat{\mathcal L}_{ij}$ on $\tilde{k}_\mu$ becomes as usual
\begin{equation}
 \hat{\mathcal L}_{ij}\tilde{k}_0=0,~~~~ \hat{\mathcal L}_{ij}\tilde{k}_l=-(\delta_{il}\tilde{k}_j-\delta_{jl}\tilde{k}_i). \label{k-Poincare-1}
\end{equation}
The Lorentz boost is somewhat difficult, and it is given by (Appendix B)
\begin{align}
 \hat{\mathcal L}_{i0}\tilde{k}_0 & =-\tilde{k}_i , \\
 \hat{\mathcal L}_{i0}\tilde{k}_j  & =-\left\{\delta_{ij}\left( \frac{\tilde{\boldsymbol{k}}^2}{2}+\frac{1-e^{-2\tilde{k}_0}}{2} \right) -\tilde{k}_i\tilde{k}_j \right\}. \label{k-Poincare-2}
\end{align} \\
The closed algebra (\ref{k-Poincare-1})-(\ref{k-Poincare-2}) consisting of $(\hat{\mathcal L}_{ij},\hat{\mathcal L}_{0i},k_\mu)$ is known as the $\kappa$-Poincare algebra\cite{kappa-Minkowski}, in which by definition, $C_1=P_4(k)$ and $C_2=P(k)^\mu P(k)_\mu$ form respectively the first and the second Casimir invariants. The invariant wave equation under the $\kappa$-Poincare transformations is, thus, $P_4(k)\Psi=0$ or $P(k)^\mu P(k)_\mu\Psi=0$. We may read those equations as non-local field equations in the Minkowski spacetime by substitution $k_\mu \rightarrow i\frac{\partial}{\partial x^\mu}$. Then, those equations describe non-local fields with timelike extension, which spoils the symmetry under the Lorentz boost. It is, however, likely that if we start with a higher-dimensional spacetime with a timelike extra-dimension, then we can realize the non-commutativity between the usual spacetime and the extra-dimension, so that the Lorentz covariance is maintained. 

\section{Space non-commutatively realized in an $AdS_{n+1}$ spacetime}

We are interested in the $(n+1)$-dimensional noncommutative spacetime with the coordinates $(\hat{x}^{\hat{\mu}},\hat{x}^n)$ characterized by
\begin{align}
 [\hat{x}_{\hat{\mu}},\hat{x}_{\hat{\nu}}] &=0 , \label{al-1}\\
 [\hat{x}_{n},\hat{x}_{\hat{\mu}}] &=i\kappa^{-1}\hat{x}_{\hat{\mu}} , \label{al-2}
\end{align}
where $\hat{\mu}=(\mu,i)$ runs over $(\mu)=(0,1,2,3)$ and $(i)=(4,5,\cdots,n-1)$. The metric, here, is assumed to be $g_{\hat{\mu}\hat{\nu}}={\rm diag}(+,-,-,\cdots,--)$. One can realize the closed algebra (\ref{al-1}) and (\ref{al-2}) by the combination of the generators of isometry group of $AdS_{n+1}$ with coordinates $(y^A)=(y^{\hat{\mu}},y^n,y^{n+1})$ defined by
\begin{equation}
 g_{AB}y^A y^B=\eta_{\hat{\mu}\hat{\nu}}y^{\hat{\mu}}y^{\hat{\nu}}-(y^n)^2+(y^{n+1})^2 = \kappa^{2}. \label{AdS}
\end{equation}
In terms of those coordinates, the generators of isometry group, the $SO(2,n)$, can be written as $ M_{AB}=i\kappa^{-1}(y_A\frac{\partial}{\partial y^B}-y_B\frac{\partial}{\partial y^A}),(A,B=\hat{\mu},n,n+1)$, to which the same type of algebra as (\ref{SO(1,4)}) holds. The light-cone variables in this case is defined by $y^{\pm}=y^{n+1}\pm y^n$, by which the invariant length (\ref{AdS}) for $y=(y^{\hat{\mu}},y^{+},y^{-})$ can be written as $\bar{g}_{AB}y^A y^B=g_{\hat{\mu}\hat{\nu}}y^{\hat{\mu}}y^{\hat{\nu}}+y^{+}y^{-}$  with the metric
\begin{equation}
 (\bar{g}_{AB})=\begin{pmatrix}
1 & 0 & \cdots & 0 & 0 & 0\\
0 & -1 & \cdots & 0 & 0 & 0 \\
0 & \vdots & \ddots & \vdots & \vdots & \vdots \\
0 & \vdots & \cdots  & -1 & 0 & 0\\
0 & 0 & \cdots & 0 & 0 &\frac{1}{2} \\
0 & 0 & \cdots & 0 & \frac{1}{2} & 0
\end{pmatrix} .
\end{equation}
Then, it is easy to verify that the combination
\begin{align}
 \hat{x}_{\hat{\mu}} &={M}_{\hat{\mu},n}+{M}_{\hat{\mu},n+1}=2M_{\hat{\mu}+}, \\
\hat{x}_n &={M}_{n,n+1}=-2M_{-+},
\end{align}
satisfy Eqs.(\ref{al-1}) and (\ref{al-2}). 

As in the previous section, we can again construct the vector in $AdS_{n+1}$ space using the contracted $SO(2,n)$ transformation $U(\omega)=e^{\frac{1}{2}\omega^{BC}M_{BC}}, (\omega^{BC}=a^Bb^C-b^Ba^C)$ associated with the light-like vector $b=(b^0,\cdots,b^{n-1},b^n,b^{n+1})=(0,\cdots,0,1,1)$. Then one can verify that $\frac{1}{2}\omega^{BC}M_{BC}=a^{\hat{\mu}}\hat{x}_{\hat{\mu}}-a^{-}\hat{x}_n$; and, the finite transformation defined by $y^A(\omega)=U(\omega)y^AU^\dag(\omega)=(e^{\kappa^{-1}\boldsymbol{\omega}})^A{}_By^B$ can be obtained as
\begin{equation} 
e^{\kappa^{-1}\boldsymbol{\omega}}=\boldsymbol{1}+\frac{1}{(a^{-})^2}\left\{\cosh (\kappa^{-1}a^{-})-1\right\}\boldsymbol{\omega}^2+\frac{1}{a^{-}}\sinh(\kappa^{-1}a^{-})\boldsymbol{\omega}.
\end{equation}
Then, each component of the vector  $\tilde{u}_A=\kappa^{-1}(e^{-\tilde{\boldsymbol{\omega}}})_{A,N+1}$ in $AdS_{n+1}$ becomes
\begin{align}
 \tilde{u}_{\hat{\mu}}(\omega) &=(e^{-\tilde{\boldsymbol{\omega}}})_{\hat{\mu},n+1}=\left\{\cosh(\tilde{a}^{-})-1\right\}\left(\frac{a_{\hat{\mu}}}{a^{-}}\right)-\sinh(\tilde{a}^{-})\left(\frac{a_{\hat{\mu}}}{a^{-}}\right), \\
 \tilde{u}_{n}(\omega) &=(e^{-\tilde{\boldsymbol{\omega}}})_{n,n+1}=\left\{\cosh(\tilde{a}^{-})-1\right\}\frac{a^{\hat{\mu}}a_{\hat{\mu}}}{(a^{-})^2}-\sinh(\tilde{a}^{-}), \\
 \tilde{u}_{n+1}(\omega) &=(e^{-\tilde{\boldsymbol{\omega}}})_{n+1,n+1}=\cosh(\tilde{a}^{-})-\left\{\cosh(\tilde{a}^{-})-1\right\}\frac{a^{\hat{\mu}}a_{\hat{\mu}}}{(a^{-})^2}, 
\end{align}
to which $\tilde{u}^A\tilde{u}_A=1$ is satisfied obviously.
 It is also straightforward to rewrite those components in terms of the wave numbers, associated with an ordered exponential function. In what follows, for a reason of symmetry, we consider the case of symmetric ordering such as
\begin{equation}
 \hat{e}_S(k)=e^{\frac{i}{2}k^{n}\hat{x}^{n}}e^{-ik^{\hat{\mu}}\hat{x}_{\hat{\mu}}}e^{\frac{i}{2}k^{n}\hat{x}^{n}}=e^{-ik_S^{\hat{\mu}}\hat{x}_{\hat{\mu}}+ik_S^{n}\hat{x}^n},
\end{equation}
which leads to
\begin{equation}
 k_S=(k_S^{\hat{\mu}},k_S^{n})=\left(\frac{\tilde{k}^n}{e^{\tilde{k}^n/2}-e^{-\tilde{k}^n/2}}k^{\hat{\mu}},k^n \right). 
\end{equation}
Then, the substitution $(a^{\hat{\mu}},a^{-})=(-k_S^{\hat{\mu}},k^n)$ gives rise to the expressions
\begin{align}
 \tilde{u}^{\hat{\mu}}(k) &=e^{-\frac{\tilde{k}^n}{2}}\tilde{k}^{\hat{\mu}}, \label{u-mu}\\
 \tilde{u}^n(k) &=-\frac{1}{2}\tilde{k}^{\hat{\mu}}\tilde{k}_{\hat{\mu}}+\sinh(\tilde{k}^n), \label{u-n} \\
  \tilde{u}^{n+1}(k) &=\cosh(\tilde{k}^n)-\frac{1}{2}\tilde{k}^{\hat{\mu}}\tilde{k}_{\hat{\mu}}. \label{u-n1}
\end{align}

In this case, one can again define a non-linear $(k^{\hat{\mu}},k^n)$ transformation, which  causes the linear $SO(2,n)$ transformation of $(\tilde{u}^A)$.  In particular, by the expressions (\ref{u-mu}),(\ref{u-n}) and (\ref{u-n1}), the $SO(1,n-1)$ transformation, the (n-1)-dimensional Lorentz transformation such as
\begin{align}
 {\mathcal L}_{\hat{\mu}\hat{\nu}}\tilde{u}_{\hat{\rho}} &=\tilde{u}_{\hat{\mu}}g_{\hat{\nu}\hat{\rho}}-\tilde{u}_{\hat{\nu}}g_{\hat{\mu}\hat{\rho}}, \\
 {\mathcal L}_{\hat{\mu}\hat{\nu}}\tilde{u}_{n} &={\mathcal L}_{\hat{\mu}\hat{\nu}}\tilde{u}_{n+1}=0,
\end{align}
are equivalent to
\begin{align}
 {\mathcal L}_{\hat{\mu}\hat{\nu}}\tilde{k}_{\hat{\rho}} &=\tilde{k}_{\hat{\mu}}g_{\hat{\nu}\hat{\rho}}-\tilde{k}_{\hat{\nu}}g_{\hat{\mu}\hat{\rho}}, \label{L1}\\
 {\mathcal L}_{\hat{\mu}\hat{\nu}}\tilde{k}_{n} &=0.
\end{align}
The  Lorentz boost which causes the mixing between a new time component $\tilde{u}^{n+1}$ and the spacetime components $\tilde{u}^{\hat{\mu}}$ can be again represented as a nonlinear transformation among $\{\tilde{k}^{\hat{\mu}} \}$; and, the resultant form is
\begin{align}
 {\mathcal L}_{\hat{\mu},n+1}\tilde{k}_{\hat{\nu}} &=e^{-\frac{\tilde{k}_n}{2}}\left[ -\frac{1}{2}\tilde{k}_{\hat{\mu}}\tilde{k}_{\hat{\nu}}+g_{\hat{\mu}\hat{\nu}}\left\{\frac{1}{2}\tilde{k}^2-\cosh(\tilde{k}_n)\right\}\right], \label{L2} \\
 {\mathcal L}_{\hat{\mu},n+1}\tilde{k}_{n} &=e^{-\frac{\tilde{k}_n}{2}}\tilde{k}_{\hat{\mu}}. \label{L3}
\end{align}
As in the case of previous section, $P_A=\kappa \tilde{u}_A$ is a momentum vector in $AdS_{n+1}$ space; and, under the transformations from (\ref{L1}) to (\ref{L3}), $C_1=\tilde{u}_n(k)$ and $C_2=\tilde{u}^{\hat{\mu}}(k)\tilde{u}_{\hat{\mu}}(k)+(P_{n+1}(k))^2$ are the first and the second Casimir invariants, respectively.

\section{Non-local field in the background of noncommutative spacetime}

Let us, now, consider the wave equation for a scalar field, which is invariant under the  $SO(1,n-1)$ transformation in $\{\tilde{u}_A \}$ space. It is obvious that the linear combinations of the first and the second Casimir invariants are those candidates, which tend to the Klein-Gordon equation in the limit $\kappa \rightarrow \infty$. In what follows, we consider a wave equation with the first Casimir invariant only because of its simple structure; that is, we put
\begin{align}
 (-2\tilde{u}_n-\tilde{m}^2)\Phi=\left[\tilde{k}^{\hat{\mu}}\tilde{k}_{\hat{\mu}}-2\sinh(\tilde{k}^n) -\tilde{m}^2\right] \Phi=0 \label{eq-0},
\end{align}
as the free field equation. Here, $m=\kappa\tilde{m}$ is a $\kappa$-dependent mass-dimension parameter that is introduced in the meantime to adjust the lowest mass for this field.

In this stage, the dimensional parameter in the theory other than the additional $m_0$ is $\kappa$ only, which characterizes the spacetime in the Planck scale physics.  We are, now, intended to modify the above field equation by introducing a new energy scale $\mu (<\kappa)$ according to the following two steps:  In the first, we note that the $k^n$ is nothing but the $a^{-}$ in $\omega^{AB}=a^{[A}b^{B]}$, which define  the vector $\tilde{u}_A=(e^{-\tilde{\boldsymbol \omega}})_{A,n+1} \in AdS_{n+1}$. Since $b$ is a fixed vector in $AdS_{n+1}$ with $b^{+}$ component only, $a^A$ may be a vector in $AdS_{n+1}$ with a free $a^{+}$ component. Then, by shifting $a^{+} \rightarrow a^{+}+\kappa^2/a^{-}$, we can put $a^A$ at a projective boundary of $AdS_{n+1}$ such as $a^Aa_A=0$, on which $SO(1,n+1)$ acts as conformal transformation. Secondly, we break this conformal symmetry by introducing a scale parameter $\mu$ lower than $\kappa$ in such a way that
\begin{align}
 a^{+}=\mu\left\{\frac{\tilde{k}^n}{2\sinh\left(\frac{\tilde{k}^n}{2}\right) }\right\}^2\sim \mu.
\end{align}
Since, this equation gives rise to $\tilde{k}^n=-\frac{1}{\mu\kappa}k^{\hat{\mu}}k_{\hat{\mu}}=-\frac{\kappa}{\mu}\tilde{k}^{\hat{\mu}}\tilde{k}_{\hat{\mu}}$, then the field equation (\ref{eq-0}) is modified so that\cite{q-bilocal}
\begin{align}
 \left[\tilde{k}^{\hat{\mu}}\tilde{k}_{\hat{\mu}}+2\sinh(\kappa_\mu \tilde{k}^{\hat{\mu}}\tilde{k}_{\hat{\mu}}) -\tilde{m}^2\right] \Phi=0,
\end{align}
where $\kappa_\mu=\frac{\kappa}{\mu}$. Here, the scale parameter $\mu$ is introduced by hand without any principle; however, it may be  natural to read $\kappa_\mu\simeq 10^4 \sim 10^5$,  the order of unification. 

The above equation is invariant under the $n$-dimensional Lorentz transformation of $\{k^{\hat{\mu}}\}$, and tends to the Klein-Gordon equation $\{(1+2\kappa_{\mu})k^{\hat{\mu}}k_{\hat{\mu}}-m^2 \}\Psi=0$ for $|k^{\hat{\mu}}k_{\hat{\mu}}| \ll \mu\kappa$. It is, thus, convenient to deal with the free field equation by adjusting the scale so that
\begin{align}
  K(\underline{k})\Psi &=W_\kappa \left\{\tilde{k}^{\hat{\mu}}\tilde{k}_{\hat{\mu}}+2\sinh(\kappa_\mu \tilde{k}^{\hat{\mu}}\tilde{k}_{\hat{\mu}})-\tilde{m}^2\right\}\Psi=0, \label{domain}\\
 &\sim (k^{\hat{\mu}}k_{\hat{\mu}}-m_0^2)\Psi=0~(~\mbox{for}~|k^{\hat{\mu}}k_{\hat{\mu}}| \ll \mu\kappa ~),
\end{align}
where $\underline{k}=(k^{\hat{\mu}})=\kappa(\tilde{k}^{\hat{\mu}})=\kappa\underline{\tilde{k}}, \, W_\kappa=\kappa^2(1+2\kappa_\mu)^{-1},$ and $m_0^2=W_\kappa \tilde{m}^2\simeq \frac{1}{2\kappa_\mu}m^2$. Then, $m_0$ becomes a very small mass parameter when we read $m_\kappa$ as an ordinary low energy mass parameter. 

Substituting $i\partial_{\hat{\mu}}$ for $k_{\hat{\mu}}$ in $K(\underline{k})$, the free field equation (\ref{domain}) becomes nothing but a non-local one in $\{x^{\hat{\mu}}\}$ space, which is longer a noncommutative space. In a practical model, further, the space of extra dimensions $(x^i)=(x^4,x^5,\cdots, x^{n-1})$ must be compact. For example, if we require the $U(1)$ cyclicity $x^i\equiv x^i+2\pi r_0$, then $k^i$ takes the spectrum $k^i=\frac{l_i}{r_0},\,(l_i=0,\pm 1,\cdots)$, which we assume, henceforth, for sake of simplicity in addition to $r_0\sim \kappa^{-1}$.

\begin{center}
\begin{minipage}{6cm}
\includegraphics[width=4cm]{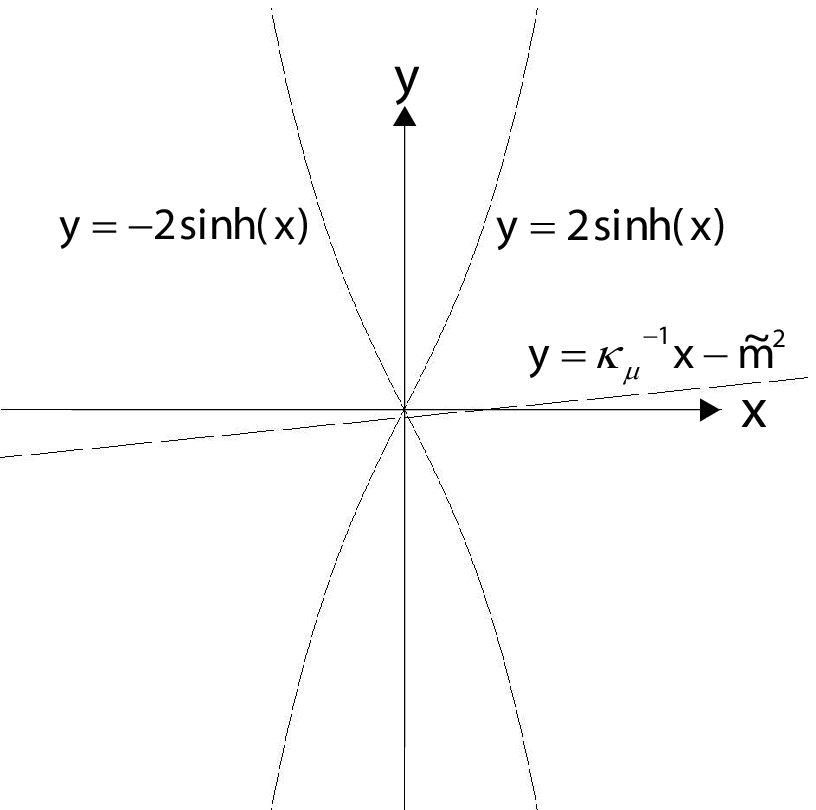}\\
{\bf Fig.1}~The intersections of $\sinh$ curves and the straight line represent the solutions of $\kappa_\mu^{-1}x \pm 2\sinh(x)-\tilde{m}^2=0$. 
\end{minipage}
\hspace{10mm}
\begin{minipage}{6cm}
\includegraphics[width=4cm]{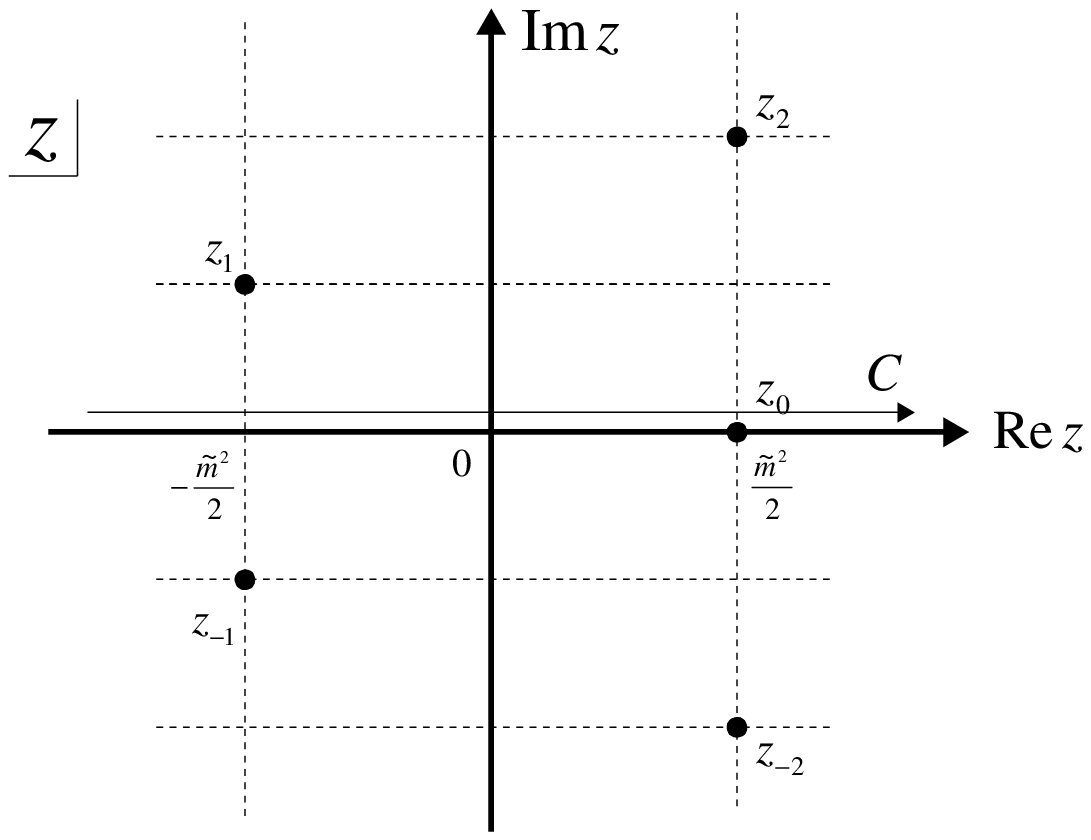}\\
{\bf Fig.2}~The solutions associated with the intersection at $x>0$ and $x<0$ in Fig.1 are corresponding to even $n$ and odd $n$ poles, respectively.
\end{minipage}
\end{center}

The solutions of $K(\underline{k})=0$, then, appear at the intersections of $y=-2\sinh(x)$ and $y=\kappa_{\mu}^{-1}x-\tilde{m}$. From Fig.1, it is obvious that the intersection gives rise to a real $\underline{k}^2>0$; i.e., a real-timelike-five momentum $\underline{k}$. Other than such a time-like solution, there are complex solutions of $\underline{k}^2$; and, the whole solutions are approximately expressed as $\underline{\tilde{k}}^2\simeq \kappa_\mu^{-1}\left\{\frac{(-1)^n}{2}\tilde{m}^2+i\pi n\right\}$, $(n=0,\pm 1,\pm 2,\cdots;|n|\lesssim 2\kappa_\mu)$
\footnote{
In terms of $z=\xi+i\eta$, the equation $\kappa_\mu^{-1}z+2\sinh(z)-\tilde{m}^2=0$ is decomposed into simultaneous equations $\kappa_\mu^{-1}\xi +2\cos(\eta)\sinh(\xi)-\tilde{m} =0$ and $\kappa_\mu^{-1}\eta+2\sin(\eta)\cosh(\xi)=0$. The latter leads to $\sin(\eta)=-(2\kappa_\mu\cosh(\xi))^{-1}\eta\simeq 0$; and so, we obtain $y\simeq \pi n,(n=0,\pm 1,\cdots; |n|\lesssim 2\kappa_\mu)$. Substituting these values for the former, the equation for $\xi$ becomes $\kappa_\mu^{-1}\xi+2(-1)^n\sinh(\xi)-\tilde{m} =0$. The Fig.1 show that the solutions for $\xi$ exist near $\xi=0$ only; and so approximating $\sinh(\xi)\simeq \xi$, we obtain $\xi\simeq \frac{(-1)^n}{2}\tilde{m}^2$. 
}.
Therefore, the mass square $M^2=k^\mu k_\mu=\bar{k}^2+k^ik^i$ of particles in 4-dimensional spacetime takes the spectra
\begin{equation}
 M^2_{n,\boldsymbol{l}} \simeq r_0^{-2}\boldsymbol{l}^2 +\frac{(-1)^n}{2} m^2+i(\mu\kappa)\pi n,~(n=0,\pm 1,\cdots; |n|\lesssim 2\kappa_\mu), \label{spectrum}
\end{equation}
where $\boldsymbol{l}=(l_4,\cdots, l_{n-1})$. 

In the right-hand side of equation (\ref{spectrum}), the first term is the order of $\kappa^2$ except the ground state $\boldsymbol{l}=0$. The third term add an imaginary component to $M^2_{n,\boldsymbol{l}}$, which may spoil the unitarity in the energy of the order of $\sqrt{\mu\kappa}$. In other words, the present effective theory will beyond the limits of validity in larger energy scale than $\sqrt{\mu\kappa}$, where the spacetime gets back to noncumulative one. We finally note that the $K^{-1}(\underline{k})$, the propagator of free field, has simple poles at 
\begin{align}
 z_n=\left\{\frac{(-1)^n}{2}\tilde{m}_\kappa^2+i\pi n\right\},~(n=0,\pm 1,\pm 2,\cdots) \label{pole}
\end{align}
 as a function of $z=\kappa_\mu\underline{\tilde{k}}^2$. Then, one can verify that
\begin{align}
 R_n\simeq \left[W_\kappa\left\{\kappa_\mu^{-1}+(-1)^n 2\cosh\left(\frac{\tilde{m}_\kappa^2}{2}\right)\right\}\right]^{-1}\simeq (-1)^n(\mu\kappa)^{-1}  \label{residue}
\end{align}
are residues of $K^{-1}(\underline{k})$ at $z=z_n$ characterized by $\left. K^{-1}\right|_{z\simeq z_n}\simeq R_n(z-z_n)^{-1}$.
Those poles are expected to play an effective role in internal lines of loop diagrams, though those poles are negligible in low energy physics.

\section{An attempt of interacting fields}

In the usual $\kappa$-Minkowski spacetime, it is not easy to formulate the interaction of fields because of its noncommutative structure among $\hat{x}^0$ and $\hat{x}^i,(i=1,2,3)$\cite{field-theory}. In our approach discussed in the previous section, the resultant spacetime is commutative one, although the fields on it obey a non-local field equation. Nevertheless, local interactions of such fields are not excluded in principle, we here study, in attempt, a $\phi^3$ type of interaction of the field, which is  characterized by the free equation (\ref{domain}) and the following action:
\begin{align}
 S[\phi]=\int d^{n+1}\left( -\frac{1}{2}\phi K(i\partial)\phi+\frac{g}{3!}\phi^3\right). \label{action}
\end{align}
For simplicity, we confine our attention to  the case of $n=5$ with the compact fifth dimension such as $x^4\equiv x^4+2\pi r_0$; and so, the wave number vector in (\ref{domain}) has the form $(k^{\hat{\mu}})=(k^\mu,r_0^{-1}l)=\kappa\{\tilde{k}^\mu,(\kappa r_0)^{-1}l \}, (l=0,\pm 1,\cdots)$.

Now, the $\sinh$ term in the free propagator $K^{-1}(\underline{k})$ plays a roll of  ultraviolet convergent for both regions of timelike and spacelike $k^{\hat{\mu}}k_{\hat{\mu}}$ in Feynman diagrams. To see this situation in detail, let us study the propagator up to the order of one-loop corrections consisting of connected diagrams described by Fig.3, to which we have the expression

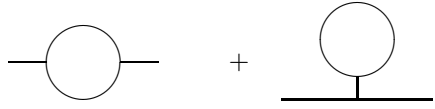
\begin{figure}
\centerline{
\begin{minipage}{3cm}
\unitlength 1mm
\begin{picture}(30,20)
\put(15,10){\circle{10}}
\put(5,10){\line(1,0){5}}
\put(20,10){\line(1,0){5}}
\end{picture}
\end{minipage}
\hspace{2mm}
$+$
\hspace{2mm}
\begin{minipage}{3cm}
\unitlength 1mm
\begin{picture}(20,20)
\put(10,13){\circle{10}}
\put(10,5){\line(0,1){3}}
\put(0,5){\line(1,0){20}}
\end{picture}
\end{minipage}}
\caption{One loop corrections to the propagator}
\label{fig:3}
\end{figure}

\begin{align}
 i\langle T(\phi_x \phi_y)\rangle_0^c &\simeq\langle x|K^{-1}(i\partial)-\frac{ig^2}{2}K^{-1}(i\partial)\left(K^{-1}*K^{-1}(i\partial)+K^{-1}(0)I\right)K^{-1}(i\partial)|y\rangle \nonumber \\
 &\simeq\frac{1}{2\pi r_0} \sum_l \int \frac{d^4p}{(2\pi)^4}\frac{e^{-ip^{\hat{\mu}}(x-y)_{\hat{\mu}}}}{K(\underline{p})-\left(\Sigma(\underline{p})+m_0^{-2}I \right)},
\end{align}
where $K^{-1}*K^{-1}$ is the convolution of $K^{-1}$. By this convolution, the self-energy term $\Sigma(\underline{p})$, the Fourier transform of $\langle x|-\frac{ig^2}{2}K^{-1}*K^{-1}(i\partial)|y\rangle$, can be expressed as
\begin{align}
 \Sigma(\underline{p}) &=-\frac{ig^2}{2}\frac{1}{2\pi r_0}\sum_l\int \frac{d^4k}{(2\pi)^4} K^{-1}(\underline{k})K^{-1}(\underline{p}+\underline{k}) . \label{self-energy}
\end{align}
Further, $I=\langle x|K^{-1}(i\partial)|x \rangle$ is the tadoploe term of Fig.4, 
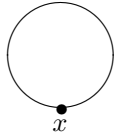
\begin{figure}
\centerline{
\begin{minipage}{4cm}
\unitlength 0.7mm
\begin{picture}(50,20)
\put(10,10){\circle{20}}
\put(9,-1.5){$\bullet$}
\put(8.5,-4.5){$x$}
\end{picture} \vspace{3mm}
\end{minipage}}
\caption{A tadpole diagram, from $x$ to $x$}
\label{fig:4}
\end{figure}
to which we have the expression
\begin{align}
 I =-\frac{ig^2}{2}\frac{1}{2\pi r_0}\sum_l \int \frac{d^4k}{(2\pi)^4}K^{-1}(\underline{k}). \label{tadopole}
\end{align}

We first evaluate the tadpole term (\ref{tadopole}) in detail, since its structure is rather simple. For this purpose, it is not available to apply a simple Wick rotation with respect to $k^0$, since $K^{-1}(\underline{k})$ has poles on complex $k^0$ plane. However, remembering that $K(\underbar{k})$ is a function of $\kappa_\mu\tilde{\underline{k}}^2=(\mu\kappa)^{-1}(k^2-r_0^{-2} l^2)$, we can write (\ref{tadopole}) in the following form:
\begin{align}
 I &=-\frac{ig^2}{2}\frac{1}{2\pi r_0}\sum_l \int \frac{d^4k}{(2\pi)^4}\int\frac{d\lambda}{2\pi}\int\frac{dz}{K[z]}e^{i\lambda(z-\kappa_\mu\underline{\tilde{k}}^2)} \nonumber \\
 &\simeq -\frac{ig^2}{2}\frac{1}{2\pi r_0(2\pi)^4}\int\frac{d\lambda}{2\pi}\left\{-\left(\frac{(\mu\kappa)\pi}{\lambda}\right)^4\right\}^{\frac{1}{2}}\int\frac{dz}{K[z]}e^{i\lambda z}, \label{tadopole-2}
\end{align}
where $K[z]=\left. K(\underline{k})\right|_{z=\kappa_\mu\underline{\tilde{k}}^2}$. We have also approximated the summation with respect to $l$ to the leading $(l=0)$ term only
\footnote{
Strictly speaking, the sum with respect to $l$ gives raise to Jacobi's theta function $\sum_l e^{-\lambda\kappa_\mu l^2}=\vartheta_3(z,q)$ with $z=1$ and $q=e^{-\lambda\kappa}$, which can be evaluated as $\vartheta_3\sim 1$ for $\lambda\gtrsim 2\kappa_\mu^{-1}$.
}
, since $l\neq 0$ makes to damp (\ref{tadopole}) by the factor $e^{-\lambda\kappa_\mu l^2}$.

The next task is to evaluate the $z$ integration in (\ref{tadopole-2}). This can be down by deforming the integration contour so as to surround poles of $K^{-1}[z]$ in (\ref{pole}) by taking their residues (\ref{residue}) into account. Then, replacing $\tilde{m}^2\rightarrow \tilde{m}^2-i\epsilon$ as usual, we obtain
\begin{align}
 \int\frac{dz}{K[z]}e^{i\lambda z} =2\pi i\left\{\theta(\lambda)\sum_{n=1}^N R_n e^{i\lambda z_n}-\theta(-\lambda)\sum_{n=0}^{-N}R_n e^{i\lambda z_n}\right\}=\frac{2\pi i}{\mu\kappa}\Theta(\lambda), 
\end{align}
where
\begin{align}
 \Theta(\lambda) \simeq \left(e^{i\lambda\frac{\tilde{m}^2}{2}}-e^{-i\lambda\frac{\tilde{m}^2}{2}+\lambda\pi} \right)\frac{1-e^{-2\pi N|\lambda|}}{e^{2\pi\lambda}-1}. \label{step-func}
\end{align}
The parameter $N(\sim 2\kappa_\mu)$ plays a role to exclude the region $\lambda \lesssim N^{-1}$. Indeed, if we approximate simply $\tilde{m}=0$ in (\ref{step-func}), we can verify that
\begin{align}
 \Theta(\lambda)\simeq 
\begin{cases}
 -\theta(-\lambda) & |\lambda| \gtrsim N^{-1} \\
 -\pi N |\lambda| & |\lambda| < N^{-1}~, \label{near zero}
\end{cases}
\end{align}
where $\theta(x)$ is the step function defined so that $\theta(x)=0$ or $1$ according as $x<0$ or $x>0$. Unfortunately, however, since the integrand of $\lambda$ integration in (\ref{tadopole-2}) has the form $\lambda^{-2}\Theta(\lambda)$, a logarithmic divergence still remains in $I$; that is, that the tadpole term $I$ can be evaluated as
\begin{align}
 I \simeq -\frac{ig^2}{2}\frac{(\mu\kappa)\pi^2}{2\pi r_0(2\pi)^4}\left[ \int_{-\infty}^{-\frac{1}{N}} \frac{d\lambda}{\lambda^2}+\pi N\int_{-\frac{1}{N}}^0 \frac{d\lambda}{|\lambda|} \right]. \label{tadopole-3}
\end{align}
The second term in the right-hand-side of the above equation is logarithmic divergent one; and so, we need a renormalization with a cut off to handle this term.

Next, let us study the self-energy term defined in (\ref{self-energy}) according to the same line of approach to $I$. By the same reason as in (\ref{tadopole}), we again discuss the case $l=0$ in both of external and internal lines in the above integral; that is, we put $\underline{p}=(p,0)$ and $\underline{k}=(k,0)$. Then, we obtain the expression
\begin{align}
 \Sigma(p,0) &\simeq -\frac{ig^2}{2}\frac{1}{2\pi r_0}\int \frac{d^4k}{(2\pi)^4} \int \! \frac{dz_1}{K[z_1]} \! \int \! \frac{dz_2}{K[z_2]}\int\frac{d\lambda_1}{2\pi}e^{i\lambda_1(z_1-\kappa_\mu\tilde{k}^2)} \! \int\frac{d\lambda_2}{2\pi}e^{i\lambda_2(z_2-\kappa_\mu(\tilde{p}+\tilde{k})^2)} \nonumber \\
 &=-\frac{ig^2}{2}\frac{1}{2\pi r_0}\frac{1}{(2\pi)^4}\int\frac{d\lambda_1}{2\pi} \! \int\frac{d\lambda_2}{2\pi}\Theta(\lambda_1)\Theta(\lambda_2) \frac{1}{i}\left[\frac{\pi(\mu\kappa)}{(\lambda_1+\lambda_2)}\right]^2e^{-i\frac{\lambda_1\lambda_2}{\lambda_1+\lambda_2}\kappa_\mu\tilde{p}^2} \nonumber \\
~. \label{sigma-1}
\end{align}
Here, since $\Theta(\lambda)=0$ for $\lambda>0$,we can insert $\int_0^\infty d\tau\delta(\tau+\lambda_1+\lambda_2)=1$ into the above integral. Then carry out the integration with respect to $\lambda_2$ after the scaling $\lambda_i=\tau\bar{\lambda}_i$, we arrive at the expression
\begin{align}
  \Sigma(p,0) &\simeq \frac{g^2}{2}\frac{1}{2\pi r_0}\frac{\pi^2}{(2\pi)^4}\int_0^\infty \frac{d\tau}{\tau} \int d\bar{\lambda}_1 \Theta\left(\tau\bar{\lambda}_1\right)\Theta\left(-\tau(\bar{\lambda}_1+1)\right)e^{i\bar{\lambda}_1(\bar{\lambda}_1+1)\tau\kappa_\mu\tilde{p}^2} \nonumber \\
 &\simeq \frac{g}{2}\frac{1}{2\pi r_0}\frac{\pi^2}{(2\pi)^4} \frac{1}{\sqrt{\kappa_\mu}} \int_0^\infty \frac{d\tau}{\tau^{\frac{3}{2}}} e^{-\frac{i\tau}{4}\kappa_\mu\tilde{p}^2}\int dx e^{ix^2\tilde{p}^2}D_\tau(x)
~, \label{sigma-2}
\end{align}
where $x=\sqrt{\tau\kappa_\mu}(\bar{\lambda}_1+\frac{1}{2})$ and
\begin{align}
 D_\tau (x)=\Theta\left(\tau(\frac{x}{\sqrt{\tau\kappa_\mu}}-\frac{1}{2})\right)\Theta\left(-\tau(\frac{x}{\sqrt{\tau\kappa_\mu}}+\frac{1}{2})\right).
\end{align}
One can find that $D_\tau (x)$ equals $1$ for almost region of $|x|<\frac{\sqrt{\tau\kappa_\mu}}{2}$ and vanishes for $|x|>\frac{\sqrt{\tau\kappa_\mu}}{2}$; that is, the interval of the integration with respect to $x$ is $-\frac{\sqrt{\tau\kappa_\mu}}{2}<x<\frac{\sqrt{\tau\kappa_\mu}}{2}$. Strictly speaking, near both limits of integration, we have to modify the edges of $D(x)_\tau$ so as to approach continuously $0$ reflecting the behavior of $\Theta(\lambda)$ near $\lambda=0$. The condition that the both ends of the interval of $x$ integration close to $0$ should be $\sqrt{\kappa_\mu\tau}\lesssim \kappa_\mu^{-1}$; that is, $\tau\lesssim \kappa_\mu^{-3}$. Under those conditions, we can put $e^{ix^2\tilde{p}^2}D_\tau(x)\simeq (\pi N)^2\tau^2\left\{\left(\frac{1}{\kappa_\mu\tau}-\frac{i\tilde{p}^2}{4}\right)x^2-\frac{1}{4}\right\}$ up to the order of $x^2$.  On the other side, we may extend the interval of $x$ integration $(-\frac{\sqrt{\kappa_\mu\tau}}{2},\frac{\sqrt{\kappa_\mu\tau}}{2})$ to all over $x$ axis for a finite $\tau \, (\gg \kappa_\mu^{-1})$. Therefore,  we can roughly evaluate the $x$ integration so that
\begin{align}
  \int dx e^{ix\tilde{p}^2}D_\tau (x) \simeq \theta(\kappa_\mu^{-3}-\tau)\frac{\pi^2}{3}\left\{(\kappa_\mu\tau)^{\frac{5}{2}}-i\tilde{p}^2(\kappa_\mu\tau)^{\frac{7}{2}}\right\}+\theta(\tau-\kappa_\mu^{-3})\sqrt{-\frac{\pi}{i\tilde{p}^2}}.
\end{align}
Substituting this expression for (\ref{sigma-2}), it follows that
\begin{align}
  \Sigma(p,0) &\simeq \frac{g^2}{2}\frac{1}{2\pi r_0}\frac{\pi^2}{(2\pi)^4} \left[ c_0+c_1\tilde{p}^2+\frac{i}{2}\sqrt{\pi\kappa_\mu} \,\, \Gamma\!\left(-\frac{1}{2},\frac{i\tilde{p}^2}{4\kappa_\mu^2}\right) \right] ~, \label{sigma-3}
\end{align}
where $c_0=\frac{\pi^2}{6}\kappa_\mu^{-4}$, $c_1=-i\frac{5\pi^2}{36}\kappa_\mu^{-6}$, and $\Gamma(-\frac{1}{2},a)$ is the incomplete gamma function with the lower limit of integration $a(=\frac{i}{4}(\tilde{p}/\kappa_\mu)^2)$, which is almost $0$ in an energy scale lower than the Planck one. Thus, the third term in the right-hand side of equation (\ref{sigma-3}) is also a constant; and, those constants in the self energy term are able to absorb into $W,\kappa,$ and $\tilde{m}^2$ in $K$. In other words, as for one-loop self-energy term, the renormalization can be carried out with finite renormalization constants.

\section{Summary and discussion}

In this paper, we have studied the $\kappa$-Minkowski spacetime from two points of view. One is the construction of  the noncommutative spacetime coordinates based on $SO(1,4)$ generators in $dS_4$ spacetime and its modification to $AdS_{n+1}$ background spacetime, which allows commutative four-dimensional spacetime. Another is a non-local field theory based on such a modified $\kappa$-Minkowski spacetime.

As for the former, in section 2, we could show that the noncommutative coordinates $(\hat{x}_0,\hat{x}_i)$ in four-dimensional $\kappa$-Minkowski spacetime are nothing but generators of transformations between light-cone coordinate $y^{+}$ and others $(y^{-},y^i)$ in $dS_4$. The plane wave in the $\kappa$-Minkowski spacetime, then, has the meaning of a finite $SO(1,4)$ transformation. From this definition of the plane wave, the five-momentum $P_A$ in $dS_4$ associated with the bi-covariant differential of the plane wave is naturally understood as a resultant vector obtained by a finite transformation of $e_4=(0,0,0,0,1)$. 

The invariant wave equations in the $\kappa$-Minkowski spacetime are defined in terms of the first or the second Casimir invariants in the background $SO(1,4)$ symmetry. Our attention is that such a wave equation defines a non-local field theory having a similarity to Yukawa's domain theory, though the wave equation spoils four-dimensional Lorentz invariance. To secure the Lorentz invariance, in section 3, we  studied a noncommutative spacetime associated with $AdS_{n+1}$ type of background spacetime. In such a spacetime, there appears another time-like coordinate $y_{n+1}$ in addition to $y_0$, from which one can construct a $\kappa$-Minkowski like spacetime characterized by the non-commutativity $[\hat{x}_n,\hat{x}_{\hat{\mu}}]=i\kappa^{-1}\hat{x}_{\hat{\mu}}$ and $[\hat{x}_{\hat{\mu}},\hat{x}_{\hat{\mu}}]=0$. In section 4, we put the wave equation in this spacetime by using the first $SO(2,n+1)$ Casimir invariant. Then, the wave equation is not invariant under the transformations between $\hat{x}_{\hat{\mu}}$ and $\hat{x}_{n}$ but is invariant under the Lorentz transformations among $\{\hat{x}_{\hat{\mu}} \}$. Further, by introducing a new scale parameter $\mu$ at the projective boundary of the $AdS_{n+1}$, the wave equation is reduced to a non-local field equation in commutative $\{x_{\hat{\mu}}\}$ spacetime, which is invariant under the Lorentz transformation.

In a resultant spacetime, we need not worry about the non-commutativity of spacetime variables; then, in section 5, we have discussed a local interaction of fields, which obeys non-local field equations characterized by a free field equation including a infinite higher derivative term such as $\sinh\{(\kappa\mu)^{-1}\partial^2\}$. There, we tried to evaluate one-loop diagrams by assuming a $\phi^3$ type of local interaction for those fields. At first, it is expected to get finite results for those diagrams, since the $\sinh$ term in the propagator plays a roll of strong dumping factor in the both spacelike and timelike regions of momentum square $\underline{k}^2=\hat{k}^{\hat{\mu}}k_{\hat{\mu}}$. However, the situation is not so simple, because the propagator contains complex poles of $\underline{k}^2$, which may spoil the unitarity of the interactions in Planck energy scale. The contribution of those poles, fortunately, again produces dumping factor to internal lines of loop diagrams: the more the number of internal lines increase, the more the dumping effect grows. Those effects are not trivial, and one can expect to get convergent results by the same mechanism in higher loop diagrams too. 

We also note that the second scale parameter $\mu$ characterizing the resultant spacetime  is introduced  by hand without enough guiding principles. The wave equation, the first Casimir invariant, then, becomes a non-local field equation that resembles Yukawa's domain one to some points. One of the purposes of domain theory is to improve the divergent problem in local field theories. Therefore, the investigation of the meaning of $\mu$ in more detail will be an interesting future problem.

\section*{Acknowledgments}
The authors wish to thank the members of the theoretical group in Nihon University for their interest in this work and comments.

\appendix

\section{Bi-covariant differential of $U(\omega)$}

We, here, derive the bi-covariant differential of $U(\omega)=e^{i\Omega},~(\Omega=\frac{1}{2}\omega^{AB}M_{AB})$, the equation (\ref{bicovariant}), which yields the relation between an ordered plane wave $\hat{e}(k)$ and the corresponding momentum $P^A(k)$. The key is to notice
\begin{align}
 \Omega=a^{\hat{\mu}}\hat{x}_{\hat{\mu}}-a^{-}\hat{x}_n=-e^{4}\cdot\boldsymbol{\omega}\cdot\hat{x}
\end{align}
for $(\boldsymbol{\omega})^{AB}=\omega^{AB}=a^Ab^B-b^Aa^B$, where $(e^4)_A={\delta^4}_A$ and $\hat{x}=(\hat{x}^0,\hat{x}^i,\hat{x}^4)$. Then, the operation of differential to $\Omega$ should be defined by $[\Omega, \mbox{d}] =-\mbox{d}\Omega=-e_4\cdot\boldsymbol{\omega}\cdot dx=-e_4\kappa\cdot\tilde{\boldsymbol{\omega}}\cdot dx$, where $dx^A=d\hat{x}^A$ are commutative quantities satisfying
\begin{align}
 [M_{AB},dx_C]=i\kappa^{-1}(g_{BC}dx_A-g_{AC}dx_B),
\end{align}
which leads to the form of $n$ times commutator
\begin{align}
 [\Omega,[\Omega,[\cdots[\Omega,-\mbox{d}\Omega],\cdots]]]=(-i)^{n-1}\kappa e_4\cdot\tilde{\boldsymbol{\omega}}^{n+1}\cdot dx.
\end{align}
Then, it is straightforward to verify
\begin{align}
 \mbox{d}U(\omega) &=e^{i\Omega}(e^{-i\Omega}\mbox{d}e^{i\Omega}) =e^{i\Omega}\sum_{n=0}^\infty \frac{(-i)^n}{n!}[\Omega,[\Omega,[\cdots[\Omega,\mbox{d}],\cdots]]] \nonumber \\
 &=-e^{i\Omega}i\kappa e_4\cdot(e^{-\tilde{\boldsymbol{\omega}}}-1)\cdot dx=i\kappa dx\cdot(1-e^{-\tilde{\boldsymbol{\omega}}})\cdot e_4 e^{i\Omega} \nonumber \nonumber \\
 &=-i\kappa\left\{dx_\mu \tilde{u}^\mu+dx_4(\tilde{u}^4-1)\right\}U(\omega). \label{dU}
\end{align}
The result is a non-linear realization\cite{N-L-realization} of $dS_4$ vector $\tilde{u}$ out of $\boldsymbol{\omega}(a)$. 

As for the bi-covariant differential of ordered plane wave associated with $AdS_{n+1}$ , we can follow the same way as the one in $dS_4$. In this case,
\begin{align}
 \Omega=\frac{1}{2}\omega^{AB}M_{AB}=a^{\hat{\mu}}\hat{x}_{\hat{\mu}}-a^{-}\hat{x}_n=-e_{n+1}\cdot\boldsymbol{\omega}\cdot\hat{x},
\end{align}
from which we have $-d\Omega=\kappa e_{n+1}\cdot\tilde{\boldsymbol{\omega}}\cdot dx$ and the n times commutator $[\Omega,[\Omega,\cdots[\Omega,-d\Omega]\cdots]]=(-i)^n\kappa e_{n+1}\cdot\tilde{\boldsymbol{\omega}}^{n+1}\cdot dx$. Then the counterpart of (\ref{dU}) in the present case becomes
\begin{align}
 \mbox{d}U(\omega) &=e^{i\Omega}i\kappa e_{n+1}\cdot(e^{-\tilde{\boldsymbol{\omega}}}-1)\cdot dx=i\kappa dx\cdot(1-e^{-\tilde{\boldsymbol{\omega}}})\cdot e_{n+1} e^{i\Omega} \nonumber \nonumber \\
 &=-i\kappa\left\{dx^{\hat{\mu}} \tilde{u}_{\hat{\mu}}+dx^n\tilde{u}_n+dx^{n+1}(\tilde{u}_{n+1}-1)\right\}U(\omega). 
\end{align}

\section{The boost in terms of wave number vectors}

From Eq.(\ref{SO(1,3)}), the boost for $\tilde{u}_0$ and $\tilde{u}_i$ are given by ${\mathcal L}_{i0}\tilde{u}_0=-\tilde{u}_i$ and ${\mathcal L}_{i0}\tilde{u}_j=\delta_{ij}\tilde{u}_0$, so that ${\mathcal L}_{i0}(\tilde{u}_0^2-\tilde{u}_j^2)=0$ holds.  Then, the expressions  (\ref{u0}) and (\ref{ui}) associated with the plane wave $\hat{e}_R(k)$ lead to 
\begin{align}
 {\mathcal L}_{i0}\tilde{u}_j=-\left[({\mathcal L}_{i0}\tilde{k}_0)e^{\tilde{k}_0}\tilde{k}_j+e^{\tilde{k}_0}({\mathcal L}_{i0}\tilde{k}_j) \right]
 =-\delta_{ij}\left(e^{\tilde{k}_0}\frac{\tilde{\boldsymbol{k}}^2}{2}+\sinh(\tilde{k}_0) \right) . \label{A0}
\end{align}
Multiplying this equation by $e^{-\tilde{k}^0}\tilde{k}_j$ and summing up with respect to $j$, we obtain
\begin{align}
({\mathcal L}_{i0}\tilde{k}_0)\tilde{\boldsymbol{k}}^2+({\mathcal L}_{i0}\tilde{k}_j)\tilde{k}_j =\tilde{k}_i\left( \frac{\tilde{\boldsymbol{k}}^2}{2}+\frac{1-e^{-2\tilde{k}_0}}{2} \right)  \label{A1}
\end{align}
Similarly, the action of ${\mathcal L}_{i0}$ on $\tilde{u}_0$ yields 
\begin{align}
 e^{-\tilde{k}_0}{\mathcal L}_{i0}\tilde{u}_0 &=\left[ ({\mathcal L}_{i0}\tilde{k}_0)\frac{\tilde{\boldsymbol{k}}^2}{2}+({\mathcal L}_{i0}\tilde{k}_j)\tilde{k}_j+e^{-\tilde{k}_0}({\mathcal L}_{i0}\tilde{k}_0)\cosh(\tilde{k}_0) \right] \nonumber \\
 &= ({\mathcal L}_{i0}\tilde{k}_0) \left(\frac{\tilde{\boldsymbol{k}}^2}{2}+\frac{1+e^{-2\tilde{k}_0}}{2}\right)+({\mathcal L}_{i0}\tilde{k}_j)\tilde{k}_j=\tilde{k}_i. \label{A2}
\end{align}
Subtracting (\ref{A2}) from (\ref{A1}), we get
\begin{align}
 ({\mathcal L}_{i0}\tilde{k}_0) \left(\frac{\tilde{\boldsymbol{k}}^2}{2}-\frac{1+e^{-2\tilde{k}_0}}{2}\right)=\tilde{k}_i \left(\frac{\tilde{\boldsymbol{k}}^2}{2}-\frac{1+e^{-2\tilde{k}_0}}{2}\right)
\end{align}
; that is, that
\begin{align}
 {\mathcal L}_{i0}\tilde{k}_0=\tilde{k}_i 
\end{align}
Substituting this result for (\ref{A0}), it can be derived that
\begin{align}
 {\mathcal L}_{i0}\tilde{k}_j=\left\{\delta_{ij}\left( \frac{\tilde{\boldsymbol{k}}^2}{2}+\frac{1-e^{-2\tilde{k}_0}}{2} \right) -\tilde{k}_i\tilde{k}_j \right\}.
\end{align}

In parallel to the above, the boosts ${\mathcal L}_{\hat{\mu},n+1}\tilde{u}_A=\tilde{u}_{\hat{\mu}}g_{n+1,A}-\tilde{u}_{n+1}g_{\hat{\mu},\hat{\nu}}$ for $SO(2,n)$ vector $\{u^A(\omega)\}$ can also be rewritten in terms of $(\tilde{k}_{\hat{\mu}},\tilde{k}_n)$ associated with the symmetric ordering. First for the component $A=\hat{\nu}$, we obtain the following as the counterpart of the equation (\ref{A1}):
\begin{align}
{\mathcal L}_{\hat{\mu},n+1}\tilde{u}_{\hat{\nu}}=\frac{1}{2}\left({\mathcal L}_{\hat{\mu},n+1}\tilde{k}_n\right)e^{\frac{1}{2}\tilde{k}_n}\tilde{k}_{\hat{\nu}}+e^{\frac{1}{2}\tilde{k}_n}\left({\mathcal L}_{\hat{\mu},n+1}\tilde{k}_{\nu}\right)=g_{\hat{\mu}\hat{\nu}}\left\{\frac{1}{2}\tilde{k}^2-\cosh(\tilde{k}_n)\right\}, \label{A3}
\end{align}
from which follows
\begin{align}
 \frac{1}{2}\left({\mathcal L}_{\hat{\mu},n+1}\tilde{k}_n\right)\tilde{k}_{\hat{\nu}}+\left({\mathcal L}_{\hat{\mu},n+1}\tilde{k}_{\nu}\right)=\tilde{k}_{\hat{\mu}}e^{-\frac{1}{2}\tilde{k}_n}\left\{\frac{1}{2}\tilde{k}^2-\cosh(\tilde{k}_n)\right\}. \label{A4}
\end{align}
Secondly, ${\mathcal L}_{\hat{\mu},n+1}\tilde{u}_{n+1}=\tilde{u}_{\hat{\mu}}$ can be read as
\begin{align}
 {\mathcal L}_{\hat{\mu},n+1}\tilde{u}_{n+1}=-\left({\mathcal L}_{\hat{\mu},n+1}\tilde{k}_{\hat{\nu}}\right)\tilde{k}^{\hat{\nu}}+\left({\mathcal L}_{\hat{\mu},n+1}\tilde{k}_{n}\right)+\sinh(\tilde{k}_n)=e^{\frac{1}{2}\tilde{k}_n}\tilde{k}_{\hat{\mu}}. \label{A5}
\end{align}
Addition (\ref{A4}) to (\ref{A5}), then, yields
\begin{align}
 {\mathcal L}_{\hat{\mu},n+1}\tilde{k}_{n}=\tilde{k}_{\hat{\mu}}e^{-\frac{1}{2}\tilde{k}_n}.
\end{align}
Substituting this result for (\ref{A3}), we arrive at
\begin{align}
 {\mathcal L}_{\hat{\mu},n+1}\tilde{k}_{\hat{\nu}}=-e^{-\frac{1}{2}\tilde{k}_n}\frac{1}{2}\tilde{k}_{\hat{\mu}}\tilde{k}_{\hat{\nu}}-g_{\hat{\mu}\hat{\nu}}e^{-\frac{1}{2}\tilde{k}_n}\left\{\frac{1}{2}\tilde{k}^2-\cosh(\tilde{k})\right\}.
\end{align}
The above results are nothing but (\ref{L2}) and (\ref{L3}).

\end{document}